\documentclass[11pt, a4paper]{article}
\usepackage[utf8]{inputenc}
\usepackage[english]{babel}
\usepackage{graphicx}
\usepackage{url}
\usepackage{amsmath}
\usepackage{amssymb}
\usepackage{authblk}
\usepackage[margin=1in]{geometry}
\usepackage{hyperref}
\usepackage{xcolor}
\usepackage{tikz}
\usetikzlibrary{shapes.geometric, arrows.meta, positioning}

\newcommand{\property}[1]{\texttt{#1}}

\hypersetup{
    colorlinks=true,
    linkcolor=blue,
    filecolor=magenta,      
    urlcolor=cyan,
    pdftitle={Intelligent Knowledge Mining Framework: Bridging AI Analysis and Trustworthy Preservation},
    pdfpagemode=FullScreen,
    }

\title{Intelligent Knowledge Mining Framework: Bridging AI Analysis and Trustworthy Preservation}
\author{Binh Vu}
\affil{Faculty of Mathematics and Computer Science, FernUniversität in Hagen, Germany \\
\texttt{binh.vu@fernuni-hagen.de}}

\date{}
\begin{document}

\maketitle

\begin{abstract}
The unprecedented proliferation of digital data presents significant challenges in access, integration, and value creation across all data-intensive sectors. Valuable information is frequently encapsulated within disparate systems, unstructured documents, and heterogeneous formats, creating silos that impede efficient utilization and collaborative decision-making. This paper introduces the Intelligent Knowledge Mining Framework (IKMF), a comprehensive conceptual model designed to bridge the critical gap between dynamic AI-driven analysis and trustworthy long-term preservation. The framework proposes a dual-stream architecture: a horizontal Mining Process that systematically transforms raw data into semantically rich, machine-actionable knowledge, and a parallel Trustworthy Archiving Stream that ensures the integrity, provenance, and computational reproducibility of these assets. By defining a blueprint for this symbiotic relationship, the paper provides a foundational model for transforming static repositories into living ecosystems that facilitate the flow of actionable intelligence from producers to consumers. This paper outlines the motivation, problem statement, and key research questions guiding the research and development of the framework, presents the underlying scientific methodology, and details its conceptual design and modeling.
\end{abstract}


\section{Introduction}
The contemporary scientific landscape is widely characterized as the fourth paradigm of discovery, an era defined by data-intensive methodologies that complement traditional theory, experimentation, and simulation \cite{hey2009fourth}. This paradigm shift is driven not only by the sheer volume of data being generated but also by its unprecedented velocity, variety, and complexity, which are the defining characteristics of Big Data \cite{laney20013d}. For researchers and knowledge workers, this proliferation of data presents both profound opportunities and significant logistical and epistemological challenges. While holding the promise of uncovering insights at a scale previously unimaginable, it also threatens to outpace our capacity to manage, process, and interpret the information we create, potentially leading to a "knowledge glut" where an abundance of data coincides with a poverty of insight. The infrastructure to support this new paradigm often significantly lags, leaving vast potential unrealized.

A primary and pervasive consequence of this infrastructural deficit is the emergence of "data silos," a phenomenon where valuable datasets become isolated within specific departmental applications, proprietary software systems, incompatible formats, or the local storage of individual researchers \cite{abraham2019bridging}. These silos represent more than a mere technical inconvenience; they are fundamental structural and cultural barriers that inhibit scientific progress. For example, a medical research institution's clinical trial records may remain entirely disconnected from its proteomics laboratory's mass spectrometry data, preventing holistic analysis. The consequences of such fragmentation are severe and far-reaching. It is a significant contributing factor to the "reproducibility crisis" affecting many scientific fields, as the inability to access original datasets makes it difficult or impossible to verify, replicate, or build upon prior work \cite{baker20161500}. Furthermore, data silos lead to an inefficient allocation of resources, as funding is often consumed by redundant data collection and duplicative research efforts \cite{inmon2002corporate}. Perhaps most critically, this fragmentation impedes innovation by preventing the cross-disciplinary syntheses that are frequently the source of the most profound scientific breakthroughs.

At the heart of this challenge lies the distinction between Data, Content, and Knowledge. Data are raw, context-free facts, consisting of discrete numbers, symbols, and signals that serve as the basic building blocks of information (e.g., a genomic sequence, a sensor reading) \cite{hey2009fourth}. Content refers to the collection of data organized and formatted for human consumption, most often in unstructured or semi-structured forms such as documents, reports, images, and videos. It is the vessel in which data and information are stored and communicated \cite{aggarwal2012mining}. Knowledge, in contrast, is a higher-level abstraction. It is information that has been processed, contextualized, and synthesized, enabling inferences, judgments, and informed decisions. It represents an understanding of the relationships between information and its applicability to a specific domain  \cite{alavi2001review}. Crucially, the management and transformation of these assets, from data, through content, to knowledge, has historically been an almost entirely manual process, relying on the intensive labor of human experts. The central problem addressed by this paper is that while our ability to generate Data and encapsulate it in Content has grown exponentially, our systems for transforming that content into verifiable, reusable, and machine-readable, actionable Knowledge have not kept pace. Therefore, this framework is designed to bridge that gap by providing a roadmap for automation, transforming the manual task of knowledge discovery into an intelligent, machine-assisted process.

To address these issues, foundational concepts from both Content Management Systems (CMS) and Knowledge Management Systems (KMS) have long been established. While a CMS provides the technological backbone for managing digital content, which includes documents, media, and other vessels of information, a KMS aims for a higher goal: the systematic management of an organization's intellectual capital \cite{alavi2001review}. The evolution of KMS can be traced through distinct generations. The first was largely technology-centric, often taking the form of sophisticated CMS platforms focused on the codification and storage of explicit knowledge. The second generation recognized the limitations of this approach and shifted focus to people, emphasizing the importance of tacit knowledge and the social networks constituting "communities of practice" \cite{masic2017evolution}. The current era, however, necessitates a third generation of KMS. This next wave must intelligently synthesize the strengths of the previous two, creating a sophisticated socio-technical system where humans and Artificial Intelligence (AI) collaborate in a virtuous cycle \cite{bhatt2001knowledge}. The system itself must evolve from a passive repository to an active partner in the research process, capable of mitigating the unsustainable cognitive load on researchers \cite{sweller1988cognitive}. The core challenge, therefore, is to design a true digital ecosystem: a robust, adaptive, and interdependent platform that supports the entire research life cycle \cite{briscoe2008digital}.

In response to these systemic challenges, this paper introduces the research and development of an Intelligent Knowledge Mining Framework. To realize this, we propose the Intelligent Knowledge Mining Ecosystem Portal, a reference architecture based on web portal technology, designed to function as a holistic digital ecosystem. The IKMF is envisioned not merely as a blueprint for an advanced data portal, but as a foundational architecture whose core contribution is the synthesis of three critical capabilities that address the entire knowledge lifecycle: intelligent Knowledge Extraction and Analysis (KEA), formal Knowledge Representation and Reasoning (KRR), and robust Knowledge Archiving and Preservation (KAP). 

To effectively implement these capabilities, it is critical to distinguish between and leverage two fundamentally different AI paradigms. The first, and most prominent today, is based on statistical models, including traditional Machine Learning (ML) and modern Transformer architectures \cite{ghahramani2015probabilistic}. These systems excel at identifying complex patterns in vast datasets but are not inherently capable of formal logical reasoning; their conclusions are probabilistic, not provably correct. The second paradigm is Symbolic AI (SAI) \cite{hogan2021knowledge}, which includes systems known as reasoners. This approach is explicitly knowledge-enabled, operating on formally structured information, such as ontologies and rulebases, to perform transparent and verifiable logical deductions. Truly transforming content into knowledge requires a framework that can leverage the strengths of both paradigms. Therefore, central to our approach is a hybrid AI framework that combines the deterministic, reproducible logic of symbolic reasoners with the power of modern Large Language Models, using a formal ontology as the bridge between them. By orchestrating these components, the IKMF provides a blueprint to dissolve data silos and create a unified, trustworthy environment for collaborative discovery. The remainder of this paper details the motivations guiding this vision, presents the conceptual architecture of the framework and its underlying scientific methodology, and discusses the outlook for creating platforms based on this model that can serve as catalysts for a more open, reproducible, and efficient mode of scientific inquiry.

\subsection{Motivation}
The research and development of the IKMF is motivated by a series of deeply interconnected requirements identified over an existing portfolio of research and development projects conducted by the Multimedia and Internet Applications research group (Lehrgebiet Multimedia und Internetanwendungen) at FernUniversität in Hagen. The contemporary scientific landscape, the fourth paradigm of discovery, is defined by data-intensive methodologies \cite{hey2009fourth}. This shift, driven by the volume, velocity, and variety of Big Data \cite{laney20013d}, presents immense opportunities but also reveals critical gaps in our data management and knowledge discovery infrastructure. Our work has consistently pushed the boundaries of this infrastructure, and each step has illuminated the need for a more holistic approach.

Our foundational work began with the EU-funded MENHIR (MENtal Health monitoring through InteRactive conversations) project, which created a successful ecosystem for managing audio files and metadata in mental health research \cite{vu2021content}. This ecosystem was realized through a bespoke, cloud-based Content and Knowledge Management Ecosystem (KM-EP) \cite{vu2020taxonomy}, which was architected with several core components, including a Media Archive for the primary audio recordings, a Digital Library for related publications, and a Taxonomy Manager for classification; an Asset Manager then combined these into unified 'scientific asset packages'. Within this platform, the extraction of knowledge from the data and content was performed using advanced audio analysis, where machine learning models processed speech signals to identify and extract acoustic features indicative of emotional states, such as pitch, formants and prosody. The extracted knowledge was then formally represented as time-stamped annotations, structured in a format compatible with the MPEG-7 standard \cite{martinez2002mpeg}, which linked segments of the audio content to data points within a dimensional emotional model (e.g., valence, arousal, dominance) \cite{vu2021content}. \textit{However, while MPEG-7 provides a machine-readable structure, it lacks the formal semantics required for logical inference, making the annotations unactionable by a reasoner. Despite its success in this focused domain, the platform's limitations highlighted several fundamental challenges that would drive our future research. A primary architectural flaw was its inability to integrate and fuse multiple, live, and heterogeneous data streams from disparate sources like sensors, confining it to a single integrated multisensory data type. It also suffered from a significant semantic gap; as a data portal, it lacked the deep understanding to automatically reason about the relationships between data points, leaving the full cognitive load of synthesis on the user. Finally, the focus on active research revealed a critical long-term durability gap in digital technology. The valuable and sensitive audio data was stored but not formally preserved, leaving it vulnerable to future obsolescence and loss of context.}

The subsequent EU-funded STOP (STop Obesity Project), a project focused on health monitoring for people with obesity, directly resolved MENHIR's fusion problem by creating a middleware platform capable of integrating live data from wearables, apps, and chatbots \cite{vu2022towards}. This was realized through the STOP KM-EP Platform, a multi-layered portal that adapted the KM-EP architecture to serve as a middleware for health data. It was designed with specific components for Sensory Data Fusion from wearables, an AI Assistant Chatbot for user interaction, and a Gamification App to encourage engagement. Within this platform, knowledge extraction from data and content was multi-modal: it involved (1) the semantic fusion of heterogeneous data streams from wearable sensors and user-reported information (e.g., diet, activity levels) and (2) emotion detection performed by an AI chatbot that analyzed the user's text-based conversations using deep learning models. The extracted knowledge from these disparate sources was then unified and structured into Knowledge Objects (KOs) within a central Knowledge Network. These KOs served as a formal representation of the user's holistic health status, combining physiological metrics with psychological insights. \textit{In solving the fusion problem, it revealed a new challenge: its analysis was superficial. The platform was an effective monitoring and visualization tool but was not designed for deep logic-based, i.e., automatic reasoning-based diagnostics. Although the platform could display the fused data, it lacked the semantic model to represent the related knowledge and reason about the interconnections, for example, how a reported meal in the app, a period of low activity from a tracker, and a chatbot conversation about mood might be related. This magnified the semantic gap, preventing deeper, automated, i.e., reasoning-based logical clinical insights and reinforcing the need for a true knowledge-based system rather than just a data integration layer.}

The EU-funded SMILE (Supporting Mental Health in Young People: Integrated Methodology for cLinical dEcisions) project was able to solve STOP's analytical challenge. As a platform for improving youth mental health management, it focused on integrating a wide variety of diagnostic data to generate personalized, evidence-based interventions \cite{cordis2023smile}. Here, knowledge extraction from data and content was achieved through a sophisticated, multi-modal approach that collected a wide range of "digital biomarkers" from adolescents interacting with a gamified platform. An explainable AI framework analyzed not only clinical data but also behavioral cues from gameplay, cognitive performance metrics, and spontaneous expressions in speech and language. The extracted insights were then structured using a formal knowledge representation based on a Common Semantic Data Model. This model was implemented as a comprehensive, standards-based ontology that formally described the complex relationships between clinical data, patient behaviours, and environmental factors, creating a rich knowledge graph for analysis and further reasoning. \textit{This leap in analytical complexity immediately created a new, critical bottleneck: the manual generation of coherent logic-based findings reports. As the system's insights became more complex, the burden on human experts to synthesize and document them became a major scalability issue. This also brought the reproducibility gap into sharp focus, as each complex, evidence-based intervention required a verifiable evidence trail.}

The GenDAI (Genomic applications for laboratory Diagnostics supported by Artificial Intelligence) project is resolving the manual reporting bottleneck. It leverages Large Language Models (LLMs) to automatically generate clinical interpretations from complex microbiome data \cite{krause2024using}. In this project, knowledge extraction from data and content is driven by advanced AI, most notably by applying Natural Language Processing (NLP) models to treat metagenomic sequences as a formal language, thereby identifying novel biomarkers linked to specific diseases. The machine-readable knowledge representation is centered around an Entity Name System (ENS), a technology that assigns unique Persistent Identifiers (PIDs) to every single entity in the diagnostic workflow. This includes the physical patient sample, the versions of the software used for analysis, the identified biomarkers, and the final clinical findings. This process creates a comprehensive and fully traceable knowledge graph of interconnected semantic entities, providing the structured, machine-readable foundation necessary to ground the LLM's automated report generation. \textit{However, this powerful automation makes the remaining systemic gaps clearer and more critical than ever. The extra problem introduced by GenDAI is the inherent opacity and trustworthiness of generative models. This new challenge supercharges the persistent gaps: an untrustworthy AI conclusion is the direct result of a weak semantic gap, as there is no formal machine-readable knowledge to ground the LLM, and a critical reproducibility gap, as the "black box" nature of the model makes tracing its logic difficult.}

Shifting from the deep, domain-specific challenges of GenDAI, the EUt+ (European University of Technology) initiative addressed the parallel challenge of strategic knowledge management at a broad institutional level. Its mission to offer joint European engineering degrees with self-designed, multi-campus curricula required a dynamic understanding of the entire research landscape. This led to the development of Research and Innovation Capacity Maps, a platform that focuses on the automatic mapping of research capabilities throughout the university alliance. Knowledge representation was centered on a formal ontology, creating a comprehensive machine-readable knowledge graph that intelligently linked researchers, projects, publications, and areas of expertise, enabling advanced functionalities such as concept-based semantic search and intelligent matching for collaboration requests. \textit{Despite its success in providing a dynamic knowledge map of who is doing what, the system still relied on human experts for strategic interpretation. The cognitive load of synthesizing insights from this complex web of activities to generate strategic reports or identify novel, non-obvious research synergies remained a significant manual effort.}

A critical reflection on this entire portfolio of our related work reveals a recurring pattern: while each project successfully solved its immediate goals and some of its predecessor's key limitations, they all pointed towards the same set of deeper, systemic challenges. These consistent, unsolved challenges crystallize into three core motivation statements.

\vspace{3mm}

\textbf{Motivation 1}:

There was no sophisticated machine-readable semantic knowledge representation layer to enable advanced reasoning and decision support. The systems in the projects could present data to a human user, but the platforms themselves possessed no deep understanding of the context or the relationships between the data points. A primary consequence of this is the emergence of "data silos," a phenomenon where valuable datasets become isolated within specific applications, preventing holistic analysis \cite{abraham2019bridging}. This limitation meant that queries were fundamentally shallow. A user could filter for "all audio files from the project," but could not ask a complex, multi-hop question that requires traversing a rich network of relationships. Answering such a query requires a formal knowledge graph representation of the domain entities and their connections \cite{hogan2021knowledge}. Consequently, the entire cognitive load of connecting disparate information and drawing new conclusions was left to the human user. 

\textbf{Motivation Statement 1 (MS1)}:
\textit{This persistent cognitive burden on the human user, and the resulting barrier to deeper insight, provides the core motivation for developing a sophisticated semantic knowledge representation layer that can transform the system from a passive repository into an active partner in discovery.}

\vspace{3mm}

\textbf{Motivation 2}:

As a direct consequence of the previous two shortcomings, the systems only partially addressed the growing crisis of research reproducibility. This crisis, affecting many scientific fields, stems in part from the inability of independent researchers to access original datasets and methods to verify, replicate, or build upon prior work \cite{baker20161500}. In our prior work, the links between a publication, the specific version of the dataset it analyzed, the exact source code of the analysis script used, and the software environment in which it was run were implicit and not formally managed by the system. 

\textbf{Motivation Statement 2 (MS2)}:
\textit{The critical need to ensure scientific integrity and enable verifiable research, thereby fully supporting the FAIR Guiding Principles \cite{wilkinson2016fair}, is the primary motivation for a system that can formally capture and model the entire chain of provenance from raw data to final publication.}

\vspace{3mm}

\textbf{Motivation 3}:

A formal, long-term digital preservation strategy was not a core, integrated component of the architecture. While data was stored, this is fundamentally different from active, long-term preservation. Digital preservation is a set of managed activities required to mitigate the inevitable risks of "digital decay." Our previous systems lacked a proactive strategy for addressing challenges like format obsolescence, where a proprietary file format becomes unreadable because the necessary software no longer exists \cite{jain2016preservation}. Furthermore, rich contextual and provenance metadata was not captured in a standardized, permanent way using established schemas like PREMIS \cite{premis2017premis}. Without a formal framework like the Open Archival Information System (OAIS) \cite{lavorato2014oais}, there is no guarantee that a dataset stored today will be authentic and usable in the future. 

\textbf{Motivation Statement 3 (MS3)}:
\textit{The significant risk of scientific knowledge and investment being lost over time to digital decay provides the motivation to architect a formal, long-term digital preservation strategy as a foundational component of the ecosystem.}

\vspace{3mm}

The confluence of these challenges, semantic ambiguity, digital fragility, and the need for reproducible science, necessitates a solution that is holistic by design. This paper introduces the IKMF as a direct response to these motivation statements. The IKMF is envisioned as a comprehensive blueprint that moves beyond simple data access to support the entire research lifecycle, from initial analysis to verifiable, reusable, and enduring scientific discovery.

\subsection{Problem Statements and Research Questions}
The motivation statements previously outlined give rise to a set of complex, interconnected research problem areas that the IKMF is designed to address. The overarching problem is the systemic inefficiency and loss of potential knowledge caused by the current state of data fragmentation and the inadequacy of existing management tools. This general problem can be decomposed into three primary domains of inquiry, each culminating in a specific research question that guides this work.

\vspace{5mm}

\textbf{Problem Description 1}: The first major challenge lies in transforming the vast and varied sources of modern data and content into a coherent, machine-readable body of logically actionable knowledge. Simply extracting features is not enough; a formal structure is needed to give these features context and meaning, enabling advanced reasoning. To find a solution, we must conduct research in the following areas:

\begin{itemize}
    \item \textbf{Problem Area 1.1:} The first step is to turn raw data into candidate knowledge concepts. Research is needed to develop and integrate robust pipelines for extracting entities (e.g., people, organizations, proteins), attributes (e.g., dates, measurements), and preliminary relationships from a wide array of sources, including unstructured text (via NER and relation extraction), tables within documents, and objects from images. These extractions form the raw material for the semantic knowledge representation layer.
    
    \item \textbf{Problem Area 1.2:} This is where the knowledge structure and semantic meaning of knowledge representation and organization are imposed. The structure must come from a formal model. Research is needed to design and maintain a formal domain ontology, in collaboration with subject matter experts, that defines the core concepts (classes like \property{Project}, \property{Person}), their properties, and the valid relationships between them. The conceptual semantics then come from the process of mapping. Research must focus on robust methods for mapping the raw knowledge concepts extracted in Problem Area 1.1 onto this formal structural knowledge, i.e., ontology. This involves entity linking (e.g., mapping the string "Dr. Smith" to an instance of the \property{Person} class) and relation mapping (e.g., mapping the phrase "is principal investigator for" to the \property{isPrincipalInvestigatorFor} object property in the ontology). This process systematically populates a structured, unified knowledge graph.
    
    \item \textbf{Problem Area 1.3:} Once the knowledge graph is populated, it becomes an active analytical tool. With this explicit knowledge structure, research is enabled to investigate methods for higher-order logical analysis that were previously impossible. This includes leveraging the graph for complex, multi-hop queries (e.g., "Find all datasets generated by projects led by researchers from Organization X"), applying logical reasoners to infer new, implicit knowledge from the explicitly asserted facts, and using graph embedding techniques for large-scale pattern detection and link prediction.
\end{itemize}

\textbf{Problem Statement 1}: \textit{There is a lack of a cohesive and scalable framework for systematically building a semantic knowledge representation layer by bridging the gap between low-level, multi-modal feature extraction and a high-level, formal knowledge model.}

\vspace{3mm}

\textbf{Research Question 1 (RQ1): How can a scalable application solution be designed, developed, and evaluated to extract knowledge concepts from heterogeneous sources, map them structurally into a formal application domain ontology to enable construction of a unified application knowledge graph, and leverage this graph for intelligent analysis and inference?}

\vspace{5mm}

\textbf{Problem Description 2:} The third challenge is to provide a technical solution to trust and, therefore, to the scientific reproducibility crisis. This requires capturing the full "research object," which is not just data and text but the entire computational workflow that connects them. To find a solution, we have to conduct research in the following areas:

\begin{itemize}
    \item \textbf{Problem Area 2.1:} A formal model is required to act as the blueprint for reproducibility. Research must focus on designing a comprehensive provenance ontology, likely extending standards like PROV-O \cite{lebo2013prov}, to formally represent every step of a computational experiment: the exact version of the input data, the version-controlled source code of the analysis script, the complete software dependency tree (e.g., Python libraries and their versions), and the specific parameters used for execution.

    \item \textbf{Problem Area 2.2:} All such types of provenance have to be captured automatically to be reliable. Research is needed to create instrumented research environments (e.g., pre-configured containers, modified Jupyter Notebook kernels) that scientists can use for their work. These environments would be designed to automatically emit detailed provenance traces, conforming to the ontology from Problem Area 3.1, as the analysis is being executed, eliminating the need for manual record-keeping.

    \item \textbf{Problem Area 2.3:} The ultimate test of reproducibility is re-execution. Research is required to build a "Reproduction Service" that uses the captured provenance as a blueprint. This service would query the knowledge graph for a specific result, retrieve its full provenance trace, fetch the archived data and the container specification of the original environment, and automatically re-instantiate the entire experiment to validate the findings or build upon them with new data.
\end{itemize}

\textbf{Problem Statement 2}: \textit{There is no integrated system that captures, models, and preserves the complete computational, semantic, and logical provenance of a scientific result, thereby preventing verifiable reproduction.}

\vspace{3mm}

\textbf{Research Question 2 (RQ2): How can a knowledge ecosystem formally model a complete knowledge mining workflow, automatically capture its provenance through instrumented environments, and leverage this captured trace to enable automated, verifiable replication of scientific results?}

\vspace{5mm}

\textbf{Problem Description 3:} The second critical challenge is ensuring the long-term viability of the digital knowledge assets being created. This requires moving beyond passive storage to an active, intelligent preservation strategy that ensures digital objects, including knowledge objects and structures, remain usable for decades. To find a solution, we must conduct research in the following areas:

\begin{itemize}
    \item \textbf{Problem Area 3.1:} An asset's long-term value depends on its context. Research must focus on designing a system that automatically aggregates metadata from multiple sources: technical metadata extracted directly from file headers (e.g., format, checksum), semantic metadata from the knowledge graph (e.g., what entities the file is about), and contextual metadata from a Current Research Information System (CRIS) \cite{decastro2011cris} (e.g., the project and grant that funded its creation). This creates a comprehensive, 360-degree view of each knowledge asset.

    \item \textbf{Problem Area 3.2:} Active preservation requires automation. Research is needed to develop a policy-driven preservation engine \cite{rajasekar2010irods}. This involves creating a machine-readable language for defining preservation policies (e.g., "all TIFF images must have a PNG migration copy created") and a technology-watch service that monitors for format obsolescence risks. These components would drive an automated workflow orchestrator that executes preservation actions \cite{lavorato2014oais} (e.g., migration, emulation) when policies are triggered.

    \item \textbf{Problem Area 3.3:} To ensure independence from any single technology stack, research must focus on the automated creation of self-describing Archival Information Packages (AIPs) compliant with the OAIS model \cite{lavorato2014oais}. These AIPs must be designed to bundle not just the data object itself, but all of its aggregated metadata from Problem Area 2.1 and the preservation policies relevant to it, ensuring it can be understood and managed by any standards-compliant repository in the future.
\end{itemize}

\textbf{Problem Statement 3}: \textit{There is an absence of an integrated framework for active digital preservation that can automatically manage the long-term health and usability of digital knowledge assets in the face of technological obsolescence.}

\vspace{3mm}

\textbf{Research Question 3 (RQ3): How can a long-term archive of data, content, and knowledge be engineered to aggregate technical, semantic, and contextual metadata, use these rich metadata to drive automated, policy-based preservation actions, and package assets into self-describing long-term archival units to ensure future usability?}

\subsection{Methodology and Approach}
\begin{figure}[h!]
    \centering
    \includegraphics[width=0.6\textwidth]{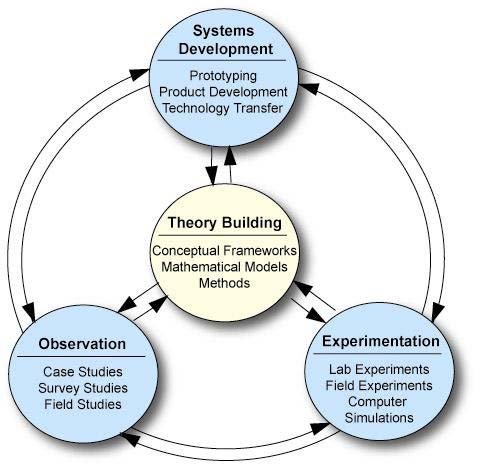}
    \caption{The Nunamaker Research Framework for Information Systems \cite{nunamaker1991systems}, illustrating the four interdependent strategies of Observation, Theory Building, Systems Development, and Experimentation.}
    \label{fig:nunamaker}
\end{figure}

The research methodology for developing and validating the IKMF is grounded in the multi-methodological framework for Information Systems (IS) research introduced by Nunamaker et al. \cite{nunamaker1991systems}. This framework, illustrated in Figure \ref{fig:nunamaker}, is a cornerstone of Design Science Research (DSR) in IS, a discipline focused on the creation and evaluation of innovative IT artifacts intended to solve real-world organizational problems \cite{hevner2004design}. The Nunamaker framework is particularly well-suited for a project of this scale and complexity as it provides a robust structure for navigating the interplay between theory and practice. It acknowledges that for complex systems, research cannot be purely theoretical or purely empirical, but must involve a constructive element where new artifacts are built and evaluated. This approach tightly couples problem identification with solution implementation and rigorous evaluation through an iterative cycle of four key research strategies \cite{hevner2004design, nunamaker2014creating}.

The framework's first component, \textit{Observation}, involves studying phenomena in their natural setting to identify and understand problems and opportunities. The second component, \textit{Theory Building}, involves creating conceptual frameworks, models, and methods that provide a formal basis for the artifact's design. The third, and often central, component is \textit{System Development}, which is the constructive phase where the theoretical model is instantiated as a tangible IT artifact, which in this research is an Ecosystem Portal. This artifact becomes a "testable theory" in its own right. The final component, \textit{Experimentation}, is used to rigorously evaluate the developed system and validate the underlying design theories. This creates a feedback loop, as the results of experimentation can lead to new observations, refinements of the theory, and further improvements to the system \cite{peffers2007design}.

\begin{figure}[h!]
    \centering
    \includegraphics[width=\textwidth]{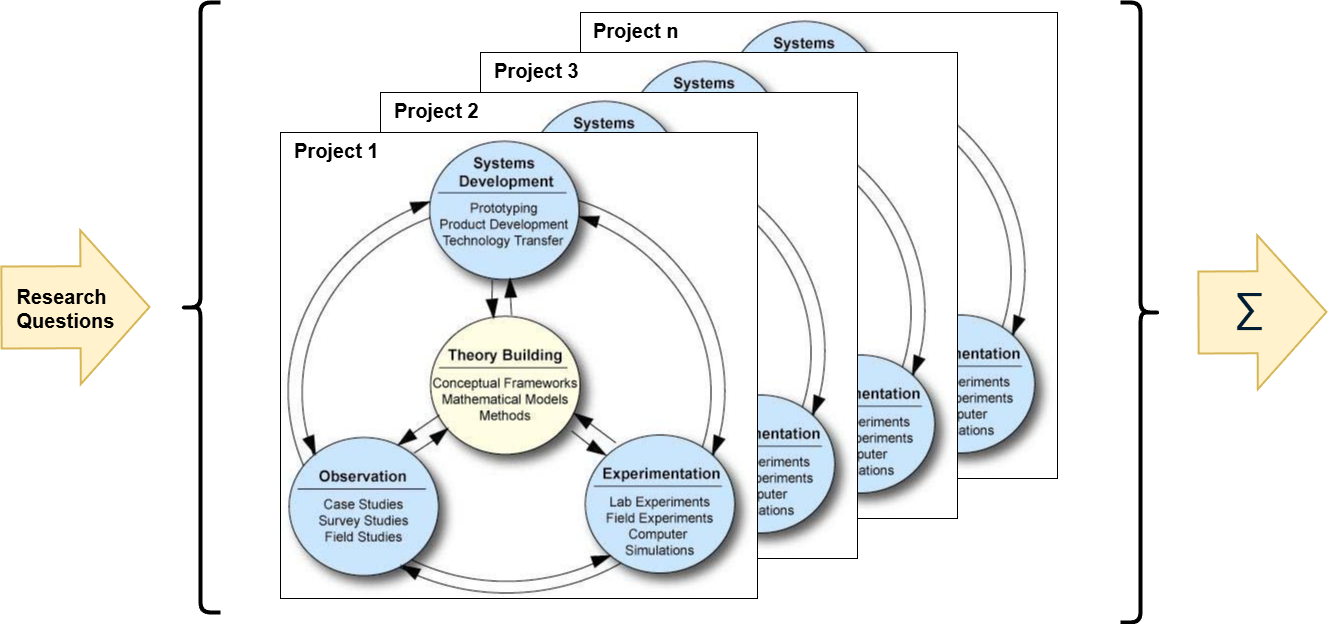}
    \caption{Decomposition of the Research Program into Targeted R\&D Projects. The overall Research Questions (RQs) are addressed through a portfolio of individual projects, each applying the full Nunamaker research cycle, with their cumulative outcomes ($\Sigma$) forming the basis for the integrated solution.}
    \label{fig:rd_projects}
\end{figure}

While this framework provides the overarching structure for the entire research program, the breadth of the three central research questions (RQ1, RQ2, RQ3) necessitates a more granular and programmatic strategy. It is impractical and methodologically unsound to attempt to address these broad, multifaceted questions within a single, monolithic research effort. Therefore, the overall research goal will be strategically decomposed into a portfolio of specific, targeted Research and Development (R\&D) projects. As illustrated conceptually in Figure \ref{fig:rd_projects}, each of these R\&D projects will, in itself, constitute a self-contained application of the Nunamaker methodology framework, focusing on a specific sub-problem of the larger vision.

This portfolio of R\&D projects will encompass a spectrum of research activities. Some projects will be foundational, focusing on basic research to advance the state-of-the-art in a specific area, such as developing a novel algorithm for semantic analysis or a new technique for data fusion under uncertainty. These projects would contribute deeply but narrowly to one of the research questions. Other projects will be more applied or integrative, concentrating on the engineering challenges of combining and adapting existing methods to work together seamlessly within the IKMF architecture. For instance, an integrative project might focus on creating a robust workflow from text extraction to knowledge graph population, drawing upon the validated outputs of more foundational projects. This strategic mix ensures that the research program simultaneously pushes theoretical boundaries while also addressing the practical challenges of building a large-scale, operational system, allowing for both deep scientific inquiry into core components and a pragmatic focus on system-level integration.

\begin{figure}[h!]
    \centering
    \includegraphics[width=\textwidth]{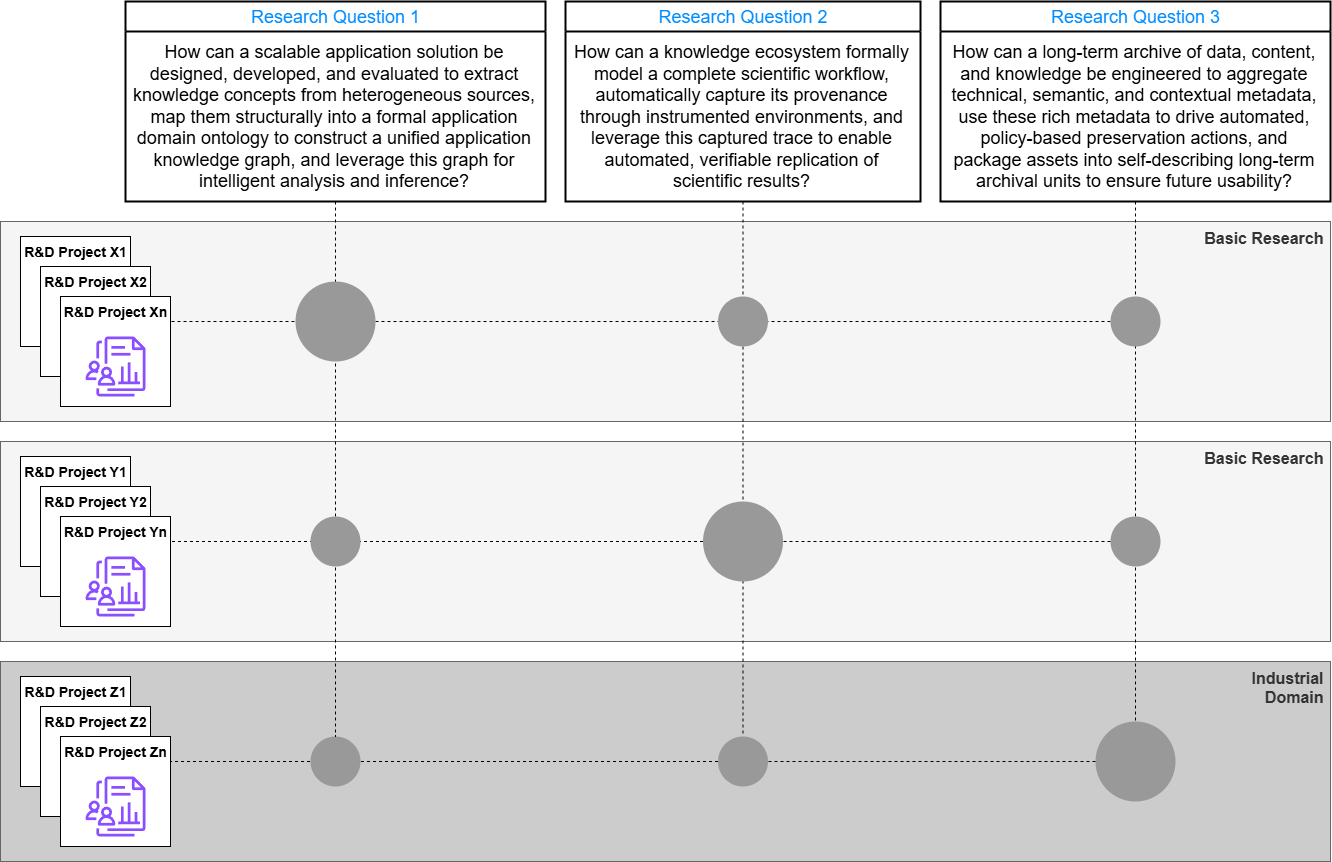}
    \caption{A Conceptual Schema for Planning and Synthesizing R\&D Project Contributions. This illustrates how different categories of projects (e.g., foundational, applied) can be strategically planned to provide weighted contributions across the three primary research questions. Adopted from \cite{marco2024}.}
    \label{fig:contribution_schema}
\end{figure}

The final and most critical step in this methodological approach is the systematic assignment and, ultimately, the synthesis of the outcomes from each R\&D project back to the overarching research questions and the integrated IKMF architecture. Figure \ref{fig:contribution_schema} provides an exemplary conceptual schema for this process. This schema is not merely an accounting tool; it is a strategic planning device. It visualizes how the portfolio of R\&D projects provides balanced coverage across the three core research areas. The outcomes of each R\&D project can be thought of as contributions, new theories, validated methods, or proven software components that address one or more of the primary research questions.

The synthesis phase itself represents a significant engineering and research challenge, focusing on the interoperability and integration of the validated components. This requires defining clear architectural interfaces, establishing standard data models, and ensuring that the outputs of one project can serve as the inputs for another. This methodological plan, from decomposition into manageable R\&D projects to the final synthesis into a unified system, ensures that the ambitious vision for the IKMF is built upon a foundation of rigorously developed, empirically validated, and architecturally coherent components. It provides a clear pathway from foundational research to the creation of a complex, high-impact information system.

The trajectory of the projects discussed, from MENHIR to GenDAI, demonstrates a clear and evolving need for a unified architectural blueprint that can holistically address these recurring challenges. To provide a consistent and scalable solution, a formal reference model is required. The primary objective of this paper, therefore, is to propose and conceptually detail the IKMF, which serves as the foundational architecture for the author's overarching research program.

\section{State of the Art in Science and Technology}
The conceptual foundation of the IKMF draws upon several mature and evolving fields of computer science. This chapter presents a structured review of the state of the art, organized to first establish a broad context before systematically addressing each of the project's core research questions. The review begins by examining the evolution of Knowledge Management Systems (KMS), which provides the overarching organizational and philosophical framework for the IKMF.

Following this contextual overview, the subsequent sections are structured to directly address each of the primary research questions in sequence. The discussion first delves into the state of the art in Knowledge Extraction and Analysis (KEA) to provide the foundation for answering RQ1. It then transitions to the field of Knowledge Representation and Reasoning (KRR) to establish the current landscape relevant to RQ2. Finally, the chapter surveys established practices in Knowledge Archiving and Preservation (KAP), which directly inform the approach to RQ3. In each of these areas, the review will identify not only the foundational technologies upon which the IKMF can be built but also the critical remaining challenges and research gaps. This systematic approach ensures that the conceptual architecture, detailed in the following chapter, is a direct and innovative response to the limitations of the current state of the art.

\subsection{Knowledge Management Systems}
The discipline of Knowledge Management (KM) provides the organizational and philosophical context for the IKMF. The evolution of KMS has progressed through distinct generations, each reflecting a more sophisticated understanding of how knowledge is created, utilized, and valued within an organization \cite{masic2017evolution}. First-generation systems, emerging in the 1990s, were primarily technology-driven and centered on a codification strategy. They aimed to capture and store explicit knowledge, information that can be easily articulated and written down, in centralized repositories like document management systems and databases. The core focus was on a "people-to-document" model, effectively treating knowledge as an object to be stored and retrieved \cite{lindgren2004evolution}. 

\begin{figure}[h!]
    \centering
    \includegraphics[width=0.6\textwidth]{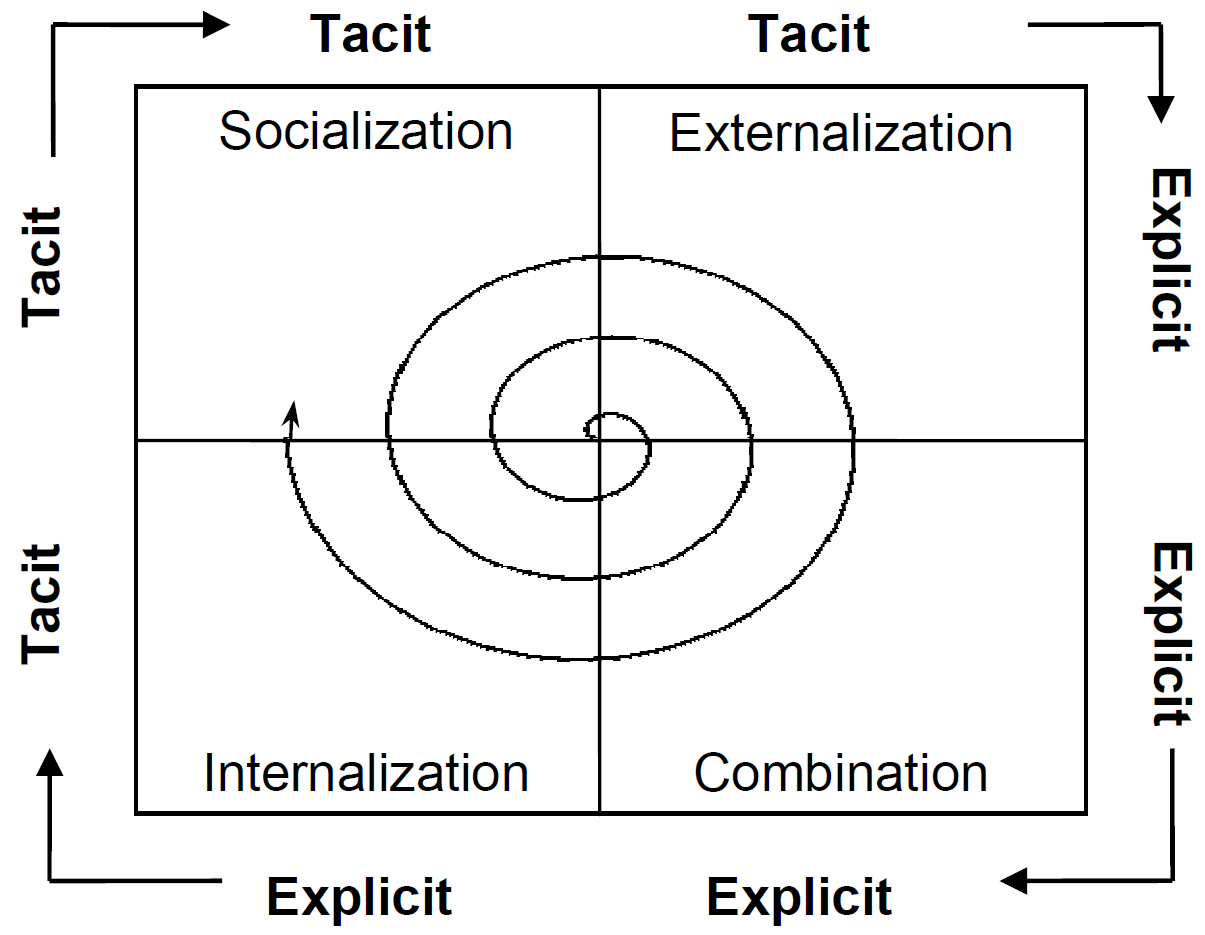}
    \caption{The SECI Model of Knowledge Creation \cite{nonaka1995knowledge}, illustrating the spiral process through which tacit and explicit knowledge are converted and amplified within an organization.}
    \label{fig:seci}
\end{figure}

A significant theoretical shift occurred with the widespread recognition of the critical importance of tacit knowledge, the unarticulated, experience-based wisdom of individuals. This led to second-generation systems, which adopted a personalization strategy focused on connecting people. These systems were heavily influenced by seminal models like Nonaka and Takeuchi's SECI process, which frames knowledge creation as a dynamic, spiraling process between tacit and explicit domains. As illustrated in Figure \ref{fig:seci}, this model posits four modes of knowledge conversion: Socialization (tacit to tacit, through shared experience and observation), Externalization (tacit to explicit, by articulating concepts and models), Combination (explicit to explicit, by synthesizing different bodies of documented knowledge), and Internalization (explicit to tacit, as individuals learn by doing and embody explicit knowledge) \cite{nonaka1995knowledge}. The focus thus shifted from documents to people, with technology being used to support "communities of practice" and connect experts with one another \cite{wenger2002cultivating}.

\begin{figure}[h!]
    \centering
    \includegraphics[width=\textwidth]{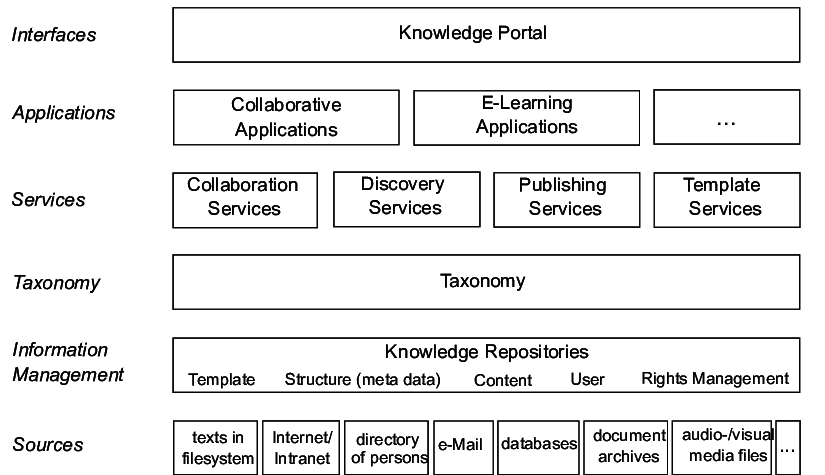}
    \caption{Layered Architecture of a Knowledge Management System. This model provides a structured view of the components required to build a comprehensive KMS, from data sources to the user-facing portal \cite{gronau2009knowledgecafe}.}
    \label{fig:gronau_arch}
\end{figure}

To bridge the gap between these high-level theories and practical implementation, structured architectural models for KMS have been proposed. A prominent example is the layered architecture described by Norbert Gronau, which has been applied in systems like "The Knowledge Café" for the hospitality industry \cite{gronau2009knowledgecafe}. As shown in Figure \ref{fig:gronau_arch}, this model provides a systematic view of a KMS, comprising several distinct, interdependent layers. At the base are the \texttt{Sources}, which include a wide variety of internal and external data in both structured and unstructured formats. This raw data is then processed, refined, and stored in \texttt{Knowledge Repositories}. A crucial organizational element is the \texttt{Taxonomy} layer, which provides the shared vocabulary and classification structure needed for consistency and semantic interoperability. On top of this infrastructure, a suite of \texttt{Services}, such as discovery, collaboration, and publishing, operate on the managed knowledge assets. These services are exposed to end-users through various \texttt{Applications}, which are finally accessed via a unified \texttt{Knowledge Portal} or interface, providing a single point of entry to the organization's collective intelligence.

Despite decades of research and development, the rate of KMS implementation failure remains alarmingly high \cite{chua2009kms}. Common reasons for failure extend beyond technical shortcomings and often lie in organizational and cultural domains. These include a lack of sustained user participation, a failure to integrate the system into existing workflows, an inability to capture the rich context surrounding information, and insufficient support from senior management \cite{akhavan2014knowledge}. Many systems fail to become a living organizational memory and instead devolve into knowledge graveyards, passive repositories of outdated and decontextualized information that users ignore. A successful KMS must overcome these hurdles. A critical factor in this success is an organization's "absorptive capacity", which is its ability to recognize the value of new information, assimilate it, and apply it effectively toward its goals \cite{cohen1990absorptive}. A KMS that is difficult to use, poorly integrated, or fails to deliver relevant insights actively hinders this capacity, preventing the organization from learning and innovating effectively.

\begin{figure}[h!]
    \centering
    \includegraphics[width=\textwidth]{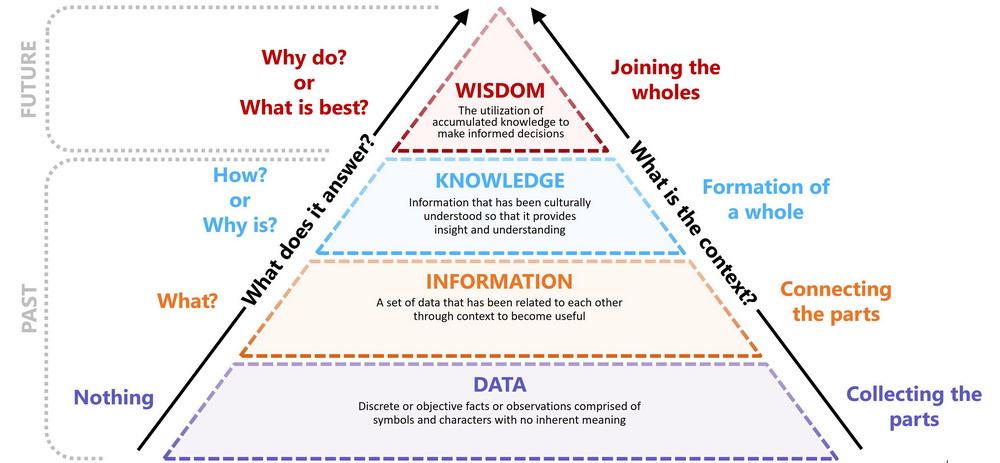}
    \caption{The DIKW Pyramid \cite{winter2023dikw}. This model illustrates the hierarchical process of transforming raw data into actionable wisdom. The IKMF aims to facilitate the transitions between each level.}
    \label{fig:dikw}
\end{figure}

The IKMF is envisioned as a third-generation KMS realized through an Ecosystem Portal reference architecture. It seeks to become an active organizational memory system, one that not only stores information but also understands its relational context through AI-driven analysis. By creating a socio-technical system where humans and AI collaborate, it aims to enhance the organization's collective absorptive capacity. The ultimate goal is to facilitate the transitions between each level of the DIKW (Data, Information, Knowledge, Wisdom) hierarchy, as illustrated conceptually in Figure \ref{fig:dikw} \cite{winter2023dikw}. In this vision, the system actively helps transform raw data into structured information, connects disparate information into contextualized knowledge, and provides the synthesized insights upon which human wisdom and novel discovery can be built.

While theoretical frameworks like the SECI model provide a powerful lens for understanding the flow of knowledge, they do not in themselves solve the problem of its preservation and utility over time. The Externalization step, where tacit knowledge is made explicit, is a critical point of fragility. Even when successfully captured in a document or database, this explicit knowledge is highly susceptible to loss, decontextualization, or staleness, which leads directly to the "knowledge graveyard" problem and the high failure rate of many KMS initiatives. This reveals a foundational challenge that underpins the entire effort: The creation of a persistent and active organizational memory. A successful system must do more than simply store the artifacts of the knowledge creation cycle, it must actively manage the lifecycle of the knowledge itself, preserving its context, ensuring its continued relevance, and preventing the inevitable decay and loss of vital intellectual capital. The technical challenges identified in the subsequent sections, which are multimodal extraction, lifecycle management, and reproducibility, are all in service of solving this fundamental problem.

\subsection{Knowledge Extraction and Analysis (KEA)}
\begin{figure}[h!]
    \centering
    \includegraphics[width=\textwidth]{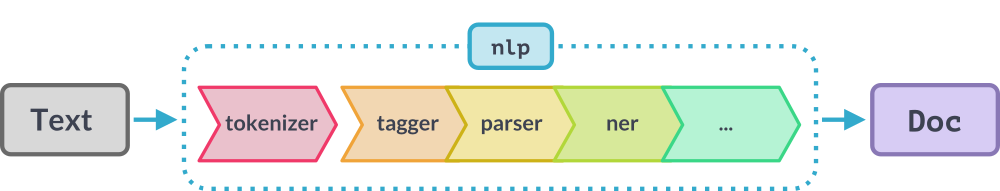}
    \caption{A conceptual NLP processing pipeline \cite{spacy2024pipeline}, illustrating how raw text is transformed into a richly annotated document (`Doc`) object through a series of modular components.}
    \label{fig:kea_pipeline}
\end{figure}

The domain of KEA is concerned with methods for identifying and extracting structured information from unstructured or semi-structured sources and subsequently analyzing this information to discover patterns. A typical KEA process begins with Information Retrieval (IR) to identify a relevant subset of documents from a large corpus. State-of-the-art IR has evolved significantly from classical sparse vector space models like TF-IDF and BM25 \cite{robertson2009probabilistic} to modern dense retrieval methods that use deep learning models to find conceptually related documents \cite{karpukhin2020dense}. Once relevant documents are retrieved, the core of knowledge extraction is performed via Information Extraction (IE) techniques, often implemented as a multi-stage NLP processing pipeline, as conceptually illustrated in Figure \ref{fig:kea_pipeline}. These pipelines modularly apply different linguistic analyses to transform raw text into annotated documents.

\begin{figure}[h!]
    \centering
    \includegraphics[width=\textwidth]{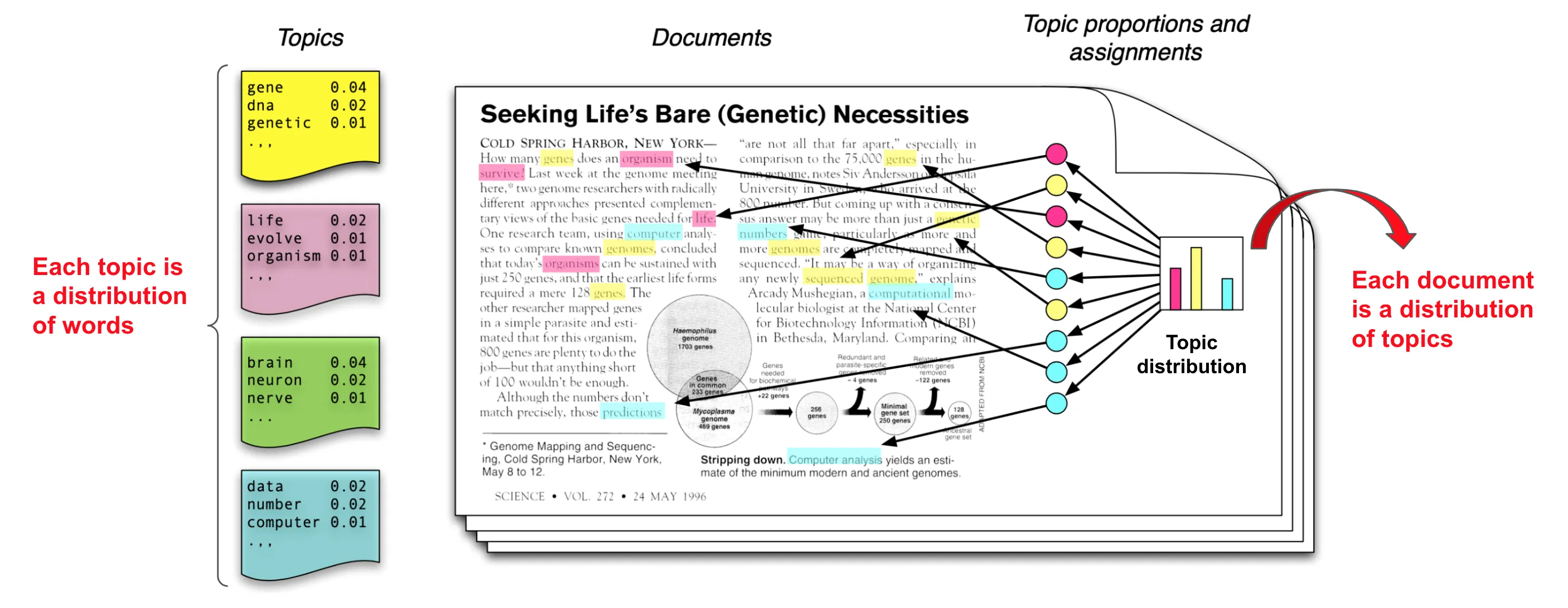}
    \caption{A conceptual illustration of the Latent Dirichlet Allocation (LDA) model. It models documents as a mixture of topics, and topics as a distribution of words \cite{aiml2024topicmodeling}.}
    \label{fig:lda_model}
\end{figure}

Beyond the foundational IE tasks of NER \cite{lample2016neural} and Relation Extraction \cite{mintz2009distant}, the state of the art in KEA includes several more advanced analytical tasks. One crucial area is Automated Text Summarization (ATS), which aims to produce a concise summary of a long document or set of documents. This is broadly divided into two approaches. Extractive summarization works by identifying and concatenating the most salient sentences from the original text, while abstractive summarization generates novel sentences, paraphrasing the source material in a manner more akin to human summarization. The latter has seen significant advances with the advent of large, pre-trained sequence-to-sequence models like BART and PEGASUS \cite{lewis2020bart, zhang2020pegasus}. Another key KEA task is Topic Modeling, an unsupervised method for discovering the latent thematic structure in a large collection of documents. The classic approach is Latent Dirichlet Allocation (LDA), a probabilistic generative model that represents documents as a mixture of topics, where each topic is a distribution over words \cite{blei2003latent}. As illustrated in Figure \ref{fig:lda_model}, LDA assumes that documents are generated by first choosing a distribution of topics for the document, and then for each word, choosing a topic and then a word from that topic's distribution.

While powerful, LDA often requires careful tuning and can struggle with short texts. More modern techniques, such as Top2Vec, leverage document and word embeddings to directly find dense clusters of documents in semantic space, often producing more coherent and interpretable topics with less manual intervention \cite{angelov2020top2vec}. Furthermore, the field of Question Answering (QA) has become a benchmark for advanced KEA. Open-domain QA systems aim to answer natural language questions by finding information within a large corpus. Early systems were extractive, trained to identify the specific text span in a document that contains the answer. Modern systems, particularly those based on large language models, are increasingly capable of generative QA, synthesizing information from multiple sources to generate a novel, abstractive answer \cite{chen2017reading}. Finally, all the information extracted from these various tasks must be consolidated. Data fusion techniques are employed to resolve entities, de-duplicate facts, and assess the veracity of information from different sources, ultimately creating a cleaner, more coherent knowledge base suitable for analysis \cite{bleiholder2008data}.

While the state of the art provides a powerful and mature toolkit for extracting structured knowledge from unstructured text, a significant gap remains. True scientific insight often requires fusing knowledge extracted not just from prose, but also from the structured data embedded within the same documents, such as tables, figures, and charts. Current pipelines often handle these different modalities in isolation, leaving the complex and often manual task of semantic fusion as a final, unresolved step. This leads to our first major research gap, which we define as Remaining Challenge 1 (RC1): The effective integration of knowledge extracted from multimodal sources. A truly intelligent system must be able to understand that a specific value in a table and a sentence in the text refer to the same entity and contribute to the same piece of knowledge.

\subsection{Knowledge Representation and Reasoning (KRR)}
\begin{figure}[h!]
    \centering
    \includegraphics[width=0.6\textwidth]{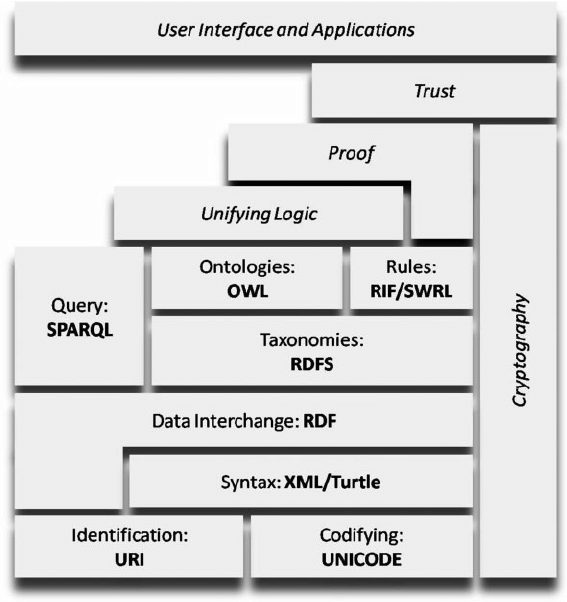}
    \caption{The Semantic Web Stack, illustrating the hierarchy of technologies from foundational syntax (XML) to logical reasoning (Proof, Trust) \cite{hoyland2014}.}
    \label{fig:semantic_web_stack}
\end{figure}

KRR is concerned with how knowledge can be formally represented so that a computer system can use it to solve complex tasks. A dominant paradigm for achieving this is the Semantic Web, a vision for a "web of data" where information is given well-defined meaning, enabling machines to understand and reason about it \cite{berners2001scientific}. This vision is realized through a stack of W3C standards, as illustrated in Figure \ref{fig:semantic_web_stack}, including the Resource Description Framework (RDF) for data modeling and the OWL for schema and logic definition \cite{w3c2014rdf, mcguinness2004owl}. Within this paradigm, various types of Knowledge Organization Systems (KOS), from simple lists to complex ontologies, are used to structure information.

\begin{figure}[h!]
    \centering
    \includegraphics[width=0.9\textwidth]{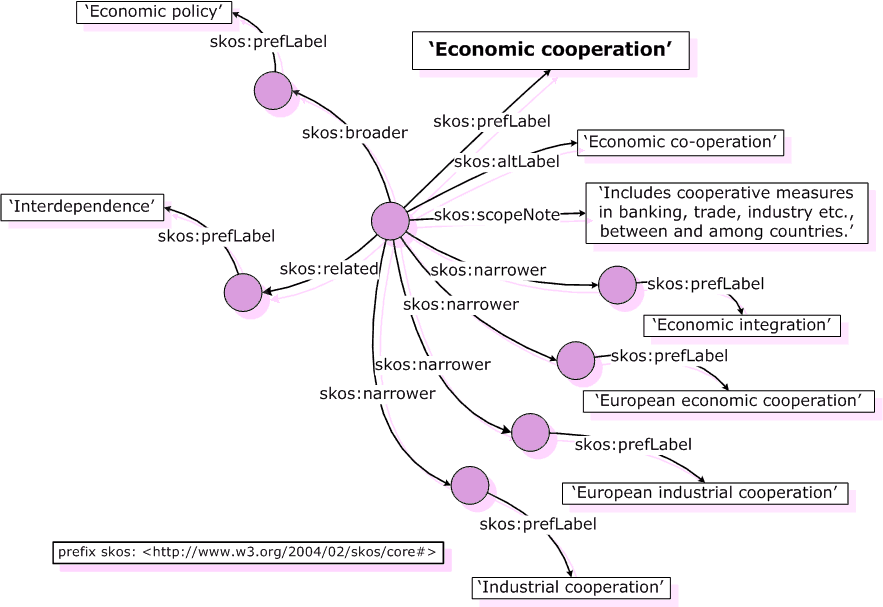}
    \caption{An example of the SKOS (Simple Knowledge Organization System) data model \cite{busse2023skos}, showing how a concept is defined through labels and linked to other concepts via hierarchical (`broader`/`narrower`) and associative (`related`) properties.}
    \label{fig:skos_model}
\end{figure}

The journey from raw data to a system capable of automated reasoning is a structured progression of increasing semantic richness. After the KEA layer has collected, cleaned, and extracted a base vocabulary of terms and named entities, the first step is to organize this vocabulary. A foundational and widely used type of KOS is the taxonomy. A taxonomy is a structured classification of concepts organized into a strict hierarchy, where each concept is a "kind of" the concept above it \cite{lambe2007organizing}. The creation and maintenance of enterprise-level taxonomies is a discipline known as Taxonomy Management \cite{hedden2016accidental}. To ensure interoperability, the W3C recommendation SKOS (Simple Knowledge Organization System) provides a standard data model for representing taxonomies in an RDF-compliant format. The core constructs of this data model are visualized in Figure \ref{fig:skos_model}. This example illustrates how a central concept, "Economic cooperation", is defined and contextualized. It is assigned a preferred lexical label (`skos:prefLabel`) and an alternative label (`skos:altLabel` as "Economic co-operation"). Its meaning is clarified with a `skos:scopeNote`. Most importantly, its position in the knowledge structure is defined through relationships: it has a broader parent concept ("Economic policy"), several narrower children concepts (e.g., "Economic integration"), and an associative, non-hierarchical link to a related concept ("Interdependence") \cite{busse2023skos}.

However, the reasoning capability within a SKOS-based taxonomy is limited. It primarily consists of "inheritance" and "transitivity." For example, if a document is tagged with "European industrial cooperation", a search for the broader term "Economic cooperation" should also retrieve this document by traversing the `skos:broader` relationships. While powerful for enhancing information retrieval and providing navigational structure \cite{gilchrist2002thesauri}, a taxonomy cannot express the rich set of relationships needed for deep reasoning. To overcome this limitation, a more expressive model is required: an ontology.

\begin{figure}[h!]
\centering
\includegraphics[width=\textwidth]{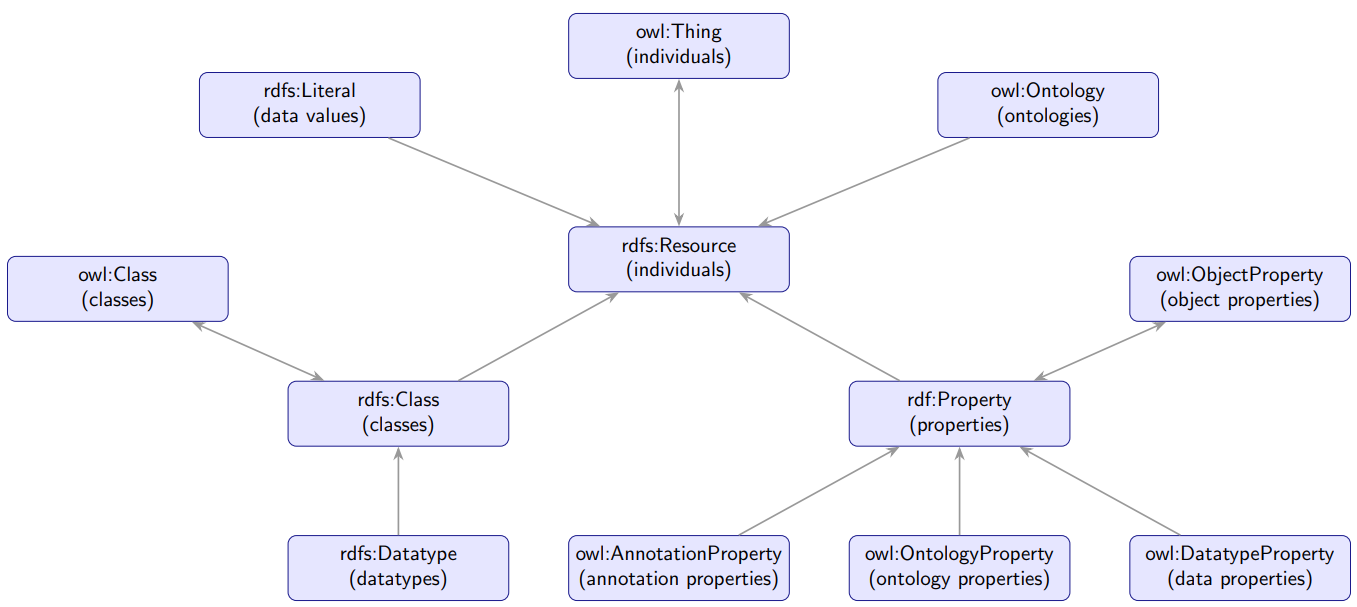}
\caption{The parts hierarchy of the OWL 2 RDF-Based Semantics. Each node represents a fundamental class (an IRI), and arrows point from a class to its superclass, showing the foundational structure of the language \cite{w3c_owl_semantics}.}
\label{fig:owl_hierarchy}
\end{figure}

An ontology, formally defined as an "explicit specification of a conceptualization" \cite{gruber1995toward}, subsumes a taxonomy but adds a formal, logical "grammar" for describing a domain. It defines classes of entities (e.g., \property{Project}, \property{Person}), the properties that describe them (e.g., \property{hasBudget}, \property{hasName}), and the different types of relationships that can exist between them (e.g., \property{isFundedBy}, \property{employs}). Crucially, an ontology includes axioms and logical constraints (e.g., a \property{Project} must be funded by an \property{Organization}). The Web Ontology Language (OWL) is the W3C standard for encoding these expressive and computationally processable ontologies \cite{mcguinness2004owl}. The foundational structure of OWL 2 is built upon the existing standards of RDF and RDFS, defining a comprehensive universe of classes and properties that serve as the building blocks for any custom ontology. This hierarchy, shown in Figure \ref{fig:owl_hierarchy}, illustrates how fundamental concepts like \property{owl:Thing} relate to RDFS resources and classes, providing the formal semantic underpinning for the language.

With a formal ontology in place, a system can perform ontological reasoning using a reasoner engine. The reasoner uses the axioms defined within the ontology to infer new knowledge and to check for logical inconsistencies. For example, if the ontology states that \property{isPrincipalInvestigatorFor} is a sub-property of \property{worksOn}, and we assert that 'Dr. Smith' \property{isPrincipalInvestigatorFor} 'Project X', the reasoner can automatically infer that 'Dr. Smith' \property{worksOn} 'Project X', even if this fact was never explicitly stated.
To capture more specific, operational logic, rule-based reasoning is added on top of the ontology. This involves defining explicit \colorbox{gray!10}{IF \textless conditions\textgreater THEN \textless conclusion\textgreater} rules. The combination of the ontology (the formal model) and the rules (the operational logic) constitutes a true Knowledge Base. Rules are the codification of expert knowledge and are primarily derived from two sources: human expertise and data-driven discovery. The most common source is domain experts, from whom a knowledge engineer elicits tacit knowledge and formalizes it into a rule language \cite{studer1998knowledge}. The second source is the data itself, where rule induction algorithms can analyze large datasets to discover statistically significant patterns that can be formulated as rules \cite{agrawal1994fast}. These rules are then defined in a formal language like the Semantic Web Rule Language (SWRL), which extends OWL to allow for the seamless integration of ontological and rule-based reasoning \cite{horrocks2004swrl}.

\begin{figure}[h!]
\centering
\includegraphics[width=0.8\textwidth]{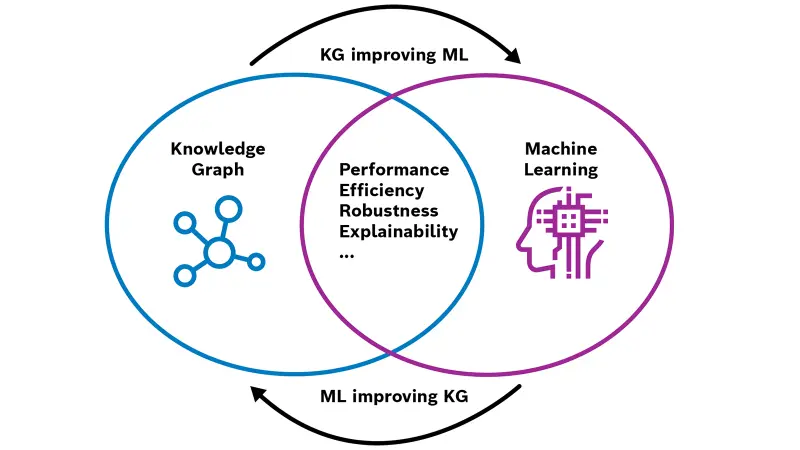}
\caption{The Neuro-Symbolic AI Cycle. Sub-symbolic models (e.g., LLMs) learn from data and propose new knowledge, while Symbolic AI (e.g., a reasoner over a Knowledge Base) provides supervision and validation, ensuring a trustworthy and continuously improving system \cite{bosch2022neurosymbolic}.}
\label{fig:neuro_symbolic_cycle}
\end{figure}

The IKMF architecture is designed not only to support "old-fashioned" or classical symbolic reasoners, but also to integrate them with new technologies like Large Language Models (LLMs) in a hybrid framework. This approach, often referred to as neurosymbolic AI, aims to combine the strengths of both paradigms to create a system that is both powerful and trustworthy \cite{garcez2019neurosymbolic}.
The symbolic reasoner remains the core engine for tasks demanding high integrity and reproducibility. Its reasoning process is deterministic and transparent; given a set of facts and rules, it will always produce the same, provably correct output. This makes it the anchor for validating the logical consistency of the knowledge graph and ensuring that any conclusions derived from it are reproducible.
In this hybrid model, the LLM acts as a powerful accelerator for knowledge extraction and a flexible natural language interface. However, its probabilistic nature and potential for factual "hallucinations" \cite{ji2023survey} make its outputs inherently non-reproducible and potentially untrustworthy. The IKMF addresses this by operationalizing the virtuous cycle of Neuro-Symbolic AI, as visualized in Figure \ref{fig:neuro_symbolic_cycle}. The subsymbolic AI component (our LLM and other ML models) learns implicit knowledge from unstructured data, discovering patterns and extracting new candidate facts. This "New Knowledge" is then used to enrich the Symbolic AI component (our Knowledge Base, comprising the ontology and rules). In the other direction, the explicit, structured knowledge from the Symbolic AI provides "Supervision" for the Sub-symbolic models. This is realized when we use the knowledge graph to ground the LLM, reducing errors and ensuring its outputs are factually consistent. This cycle of enrichment and validation allows the IKMF to leverage the speed and flexibility of modern AI while guaranteeing the logical integrity and reproducibility of its core knowledge through established symbolic methods.

While this concept is gaining traction, distinct implementation approaches exist. Frameworks such as DeepProbLog \cite{manhaeve2018deepproblog} and Scallop \cite{huang2021scallop} have successfully integrated probabilistic reasoning directly into deep learning loops to perform end-to-end differentiable reasoning. However, these are primarily algorithmic solutions focused on optimizing specific inference tasks. The IKMF distinguishes itself by applying this hybrid approach at an architectural level. Rather than focusing solely on the inference algorithm, the IKMF focuses on the lifecycle management of the ontology and the long-term preservation of the resulting inferences. It treats the neuro-symbolic interaction not just as a calculation, but as a provenance-aware process where the interaction between the LLM and the Reasoner is itself a preservable scientific record.

Despite the power of this hybrid approach, a significant practical and conceptual challenge persists in the long-term management of the knowledge base itself. Scientific understanding is not static; new discoveries are made, terminologies are refined, and relationships evolve. Most existing KRR systems treat the knowledge graph as a static endpoint of a one-time ingestion process. They often lack robust, automated mechanisms to manage schema evolution, track the provenance of every assertion, and ensure the graph remains accurate and consistent over time without costly manual intervention. This identifies our second major research gap, which we define as Remaining Challenge 2 (RC2): Managing the knowledge graph lifecycle. An effective knowledge system must treat the graph not as a static artifact, but as a living entity that can be reliably updated and maintained.

\subsection{Trustworthy Knowledge Archiving and Preservation (KAP)}
KAP is a mature discipline focused on ensuring that digital objects remain accessible, understandable, and usable over long periods of time, a critical requirement for any system intended to manage scientific knowledge. The foundational framework for this field is the OAIS reference model (ISO 14721:2012) \cite{ccsdsoais2024}. A central concept in OAIS is the 'Designated Community,' which is the specific group of consumers for whom the information is being preserved. This is not a passive audience; rather, it is the explicitly identified group whose needs, skills, and anticipated level of expertise actively shape the archive's preservation strategies. For instance, the designated community for a set of raw genomic data might be "computational biologists with expertise in bioinformatics pipelines," and all decisions about what metadata to capture and what formats to maintain would be made with their needs in mind.

\begin{figure}[h!]
    \centering
    \includegraphics[width=\textwidth]{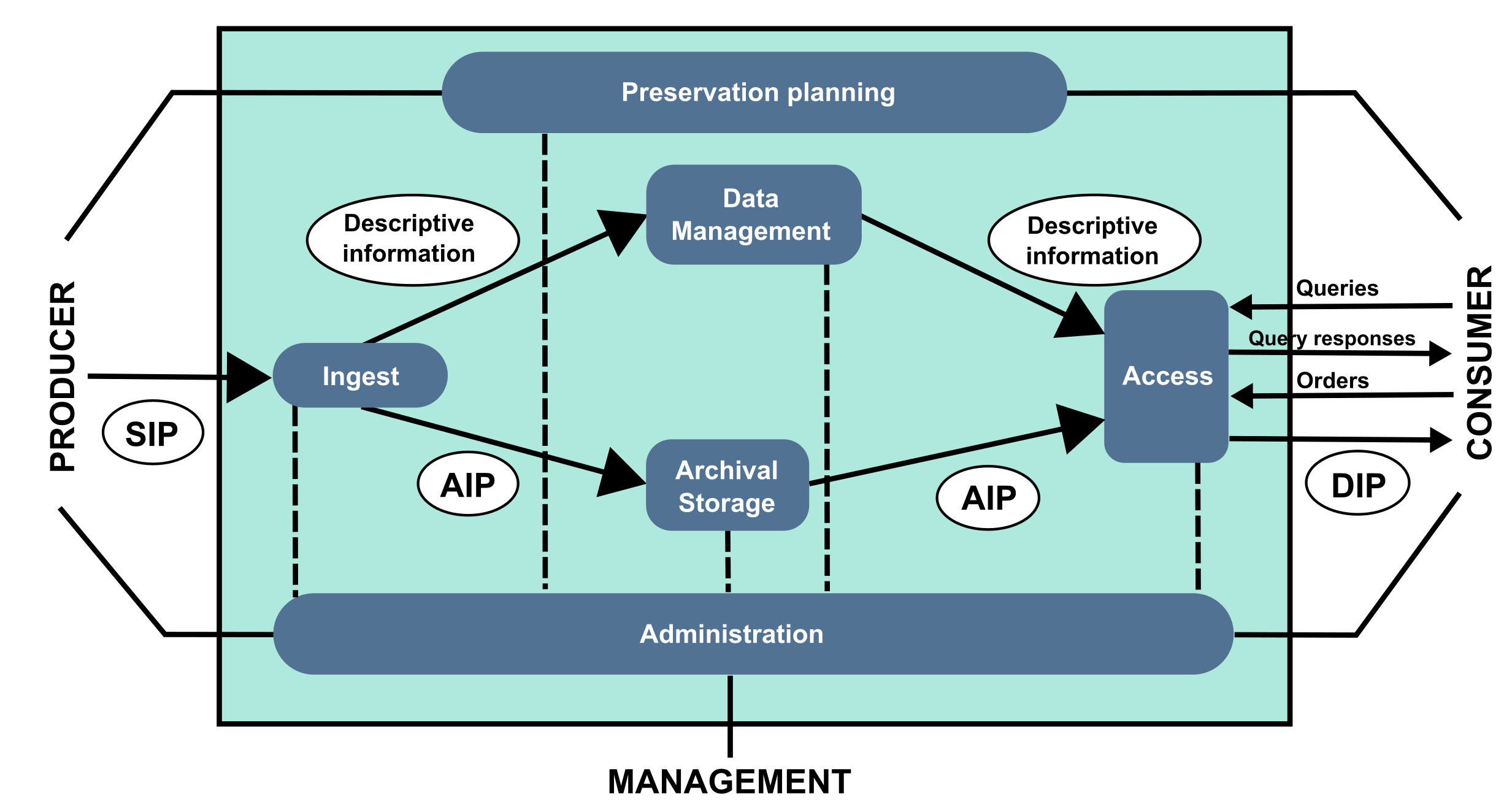}
    \caption{The Functional Model of an OAIS \cite{ccsdsoais2024}, illustrating the key entities and data flows involved in a trustworthy digital repository).}
    \label{fig:oais_model}
\end{figure}

Crucially, this Designated Community is assumed to possess a 'Knowledge Base', a set of common information, terminology, and understanding that is not explicitly included in every piece of archived information because it is considered prerequisite knowledge for that community. This Knowledge Base defines the boundary of what the archive must explicitly preserve versus what it can assume its target users already know. For example, an archive preserving data for that community of computational biologists would not need to include a definition of a FASTQ file format or the principles of DNA sequencing, as these are part of their assumed knowledge. This distinction is critical for making digital preservation practical; it prevents the archive from having to document every piece of assumed disciplinary context, focusing instead on the specific information needed to make a particular digital object independently understandable. A key innovation within the IKMF is the ability to use AI-driven knowledge extraction and representation techniques to formally model and explicitly create this Knowledge Base, transforming it from an assumed concept into a managed digital asset. As depicted in Figure \ref{fig:oais_model}, OAIS provides a comprehensive conceptual model for a digital archive, defining its key functional entities and information packages. It establishes a common vocabulary and a set of mandatory responsibilities for any organization claiming to be a trustworthy digital repository. This model is often implemented using production-ready institutional repository software, with DSpace being a prominent open-source example. DSpace is designed to capture, store, index, preserve, and redistribute digital research materials in various formats, providing the practical machinery to realize the abstract functions of the OAIS model \cite{smith2003dspace}.

\begin{figure}[h!]
\centering
\includegraphics[width=0.8\textwidth]{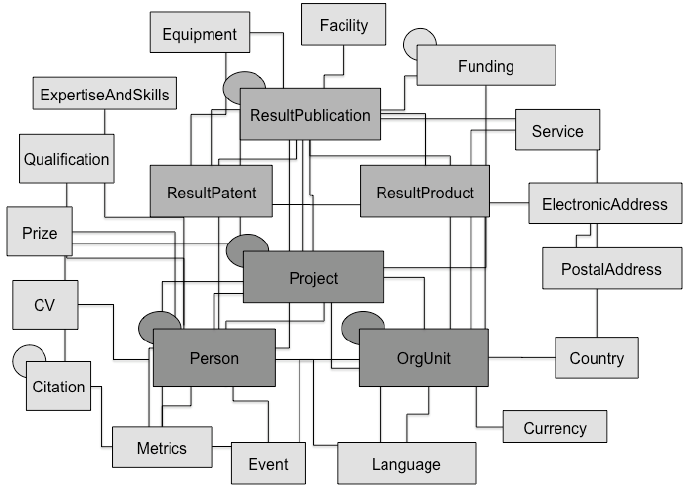}
\caption{The CERIF Data Model, illustrating how base entities (like Project and Person) are connected via semantic link entities, enabling rich, contextual relationships \cite{grootel2009}.}
\label{fig:cerif_model}
\end{figure}

A core component of any OAIS-compliant system is the requirement for robust preservation metadata. The PREMIS Data Dictionary (ISO 22927:2017) is the de facto international standard for this purpose \cite{premis2017premis}. PREMIS defines the specific information needed to support digital preservation activities, such as the technical characteristics of a file (Object), a log of all actions performed on it (Event), the people or software involved (Agent), and any associated permissions (Rights). However, for scientific knowledge, technical metadata alone is insufficient. The full context of an asset—its origin, purpose, and relationship to other research activities—is paramount for its long-term value and reusability. This is where a Current Research Information System (CRIS) becomes a critical architectural component. A CRIS is an institutional system that manages the administrative and semantic metadata about the research process itself. To standardize this contextual information and ensure interoperability between systems, a formal data model is required. The Common European Research Information Format (CERIF) provides such a standard, recommended by the European Union for its member states. As illustrated in Figure \ref{fig:cerif_model}, CERIF is a formal entity-relationship model that defines core research entities (e.g., Project, Person, Organisation, Publication, Dataset) and connects them through richly described, typed relationship entities (e.g., Person is-author-of Publication). Adopting a CERIF-compliant data model within a CRIS ensures that the research context is not only captured but is structured in a consistent, machine-readable, and exchangeable format \cite{jorg2012cerif}.

The primary technical challenge in KAP remains mitigating the risks of digital obsolescence through strategies like migration and emulation \cite{granger2000emulation}. The IKMF's KAP layer is therefore designed to implement an integrated system where the repository (DSpace) ensures technical preservation, the preservation metadata (PREMIS) guarantees authenticity and renderability, and the contextual links to the CERIF-compliant CRIS ensure that the asset's scientific meaning and provenance are permanently and interoperably maintained. Ultimately, the principles of KAP are strongly aligned with and are essential for fully realizing the FAIR Guiding Principles \cite{wilkinson2016fair}. A well-preserved digital object, enriched with its full technical and contextual metadata, is by definition more likely to remain Findable, Accessible, Interoperable, and Reusable for future generations of researchers.

To operationalize these concepts, the community has moved toward standardized packaging formats. While traditional AIPs are often repository-specific, the emergence of FAIR Digital Objects (FDOs) \cite{desmedt2020fair} provides a theoretical framework for bundling data, metadata, and persistent identifiers into machine-actionable units. In practice, standards such as RO-Crate (Research Object Crate) \cite{sefton2021rocrate} and BagIt \cite{kunze2018bagit} have gained prominence for implementing these bundles. RO-Crate, in particular, uses lightweight JSON-LD to provide semantic descriptions of the dataset and its context, directly supporting the goals of the IKMF. A key innovation of our framework is to automate the generation of these RO-Crate-compliant packages as a byproduct of the mining process, rather than requiring manual assembly by the researcher.

However, preserving the data bitstream and its metadata, while essential, is no longer sufficient to guarantee the future utility of scientific knowledge. Modern data analysis relies on complex computational workflows, involving specific software versions, their myriad dependencies, and precise execution parameters. Archiving a dataset and a publication often fails to capture the full "research object," making it impossible for another researcher to validate or replicate a result years later because the original computational environment has been lost. This critical unmet need constitutes our third major research gap, which we define as Remaining Challenge 3 (RC3): Preserving the computational reproducibility of scientific findings. 

Current solutions for this challenge include containerization technologies (e.g., Docker, Singularity) and workflow specification languages such as the Common Workflow Language (CWL) \cite{amstutz2016cwl} and Nextflow \cite{ditommaso2017nextflow}. These tools allow for the precise, machine-readable definition of execution environments and processing steps. However, they are typically utilized as manual authoring tools by advanced bioinformaticians or data engineers. There is a lack of "instrumented environments" that can automatically capture user actions in a GUI or notebook and serialize them into a CWL or Nextflow definition without explicit user intervention. A complete preservation strategy must archive not just the data, but the entire executable context that produced the knowledge from that data.

\subsection{Summary}
The state of the art in science and technology provides a strong foundation for the IKMF, yet also reveals critical gaps that this research aims to address. The evolution of Knowledge Management Systems provides the overarching context, showing a clear progression from simple document repositories to sophisticated socio-technical systems. The IKMF builds on this by proposing a third-generation, AI-driven KMS that addresses historical failure points and aims to facilitate the entire DIKW hierarchy.

The review of Knowledge Extraction and Analysis demonstrates powerful techniques for processing unstructured data, from dense retrieval methods to advanced NLP pipelines. This body of work provides a strong starting point for addressing RQ1. However, our review also surfaces a critical remaining challenge: \textit{the effective integration of knowledge extracted from multi-modal sources (RC1)}. Current systems excel at processing text but struggle to holistically fuse insights from text, tables, and figures, a gap the IKMF must bridge. For Knowledge Representation and Reasoning, the Semantic Web stack offers a robust framework for encoding knowledge, which directly informs RQ2. Despite the power of formalisms like OWL and SKOS, a significant practical challenge persists in \textit{managing the Knowledge Graph lifecycle (RC2)}. The state of the art lacks robust solutions for ensuring the knowledge base can evolve consistently as new scientific understanding emerges, a key problem for long-term knowledge management. Finally, the discipline of Knowledge Archiving and Preservation provides established models like OAIS and standards such as PREMIS, which are essential for answering RQ3. The primary remaining challenge identified is \textit{the preservation of computational reproducibility (RC3)}. Existing frameworks focus on preserving the data object, but often fail to archive the complex computational environment required to replicate the scientific findings derived from that data. Collectively, these identified gaps delineate the primary areas of innovation for the IKMF. The conceptual architecture presented in the following chapter is designed specifically to address these challenges.

\section{Conceptual Design and Modeling}
\begin{figure}[h!]
\centering
\includegraphics[width=\textwidth]{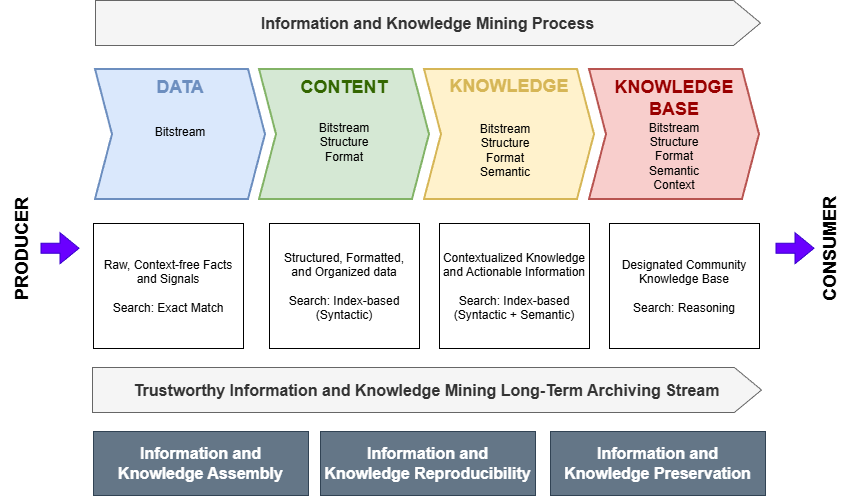}
\caption{The IKMF Reference Model, illustrating the progression from Producer to Consumer through the Information and Knowledge Mining Process, underpinned by the Trustworthy Long-Term Archiving Stream.}
\label{fig:ikmep_reference_model}
\end{figure}

Following the general structure of this research, the IKMF reference model will be outlined first. Afterwards, the integration of intelligent knowledge extraction (RQ1), formal representation and reasoning (RQ2), and trustworthy archiving (RQ3) into a unified platform for collaborative scientific discovery will be modeled and discussed. Revisiting the state of the art in knowledge management and digital preservation reveals a critical need for a system that transcends static repositories. As outlined in the previous section, the Externalization phase of the SECI model often results in "knowledge graveyards" if not actively managed. To address this, the dimensions of a functional ecosystem must move beyond simple storage to active lifecycle management. Just as trustworthiness in AI relies on dimensions like reproducibility and validity, a robust knowledge ecosystem relies on the dimensions of Semantic Enrichment, Contextual Integration, and Trustworthy Archiving. Figure \ref{fig:ikmep_reference_model} illustrates the resulting IKMF Reference Model, which provides a comprehensive blueprint for the flow of value from data producers to data consumers.

As outlined in this conceptual schematic, the architecture is defined by two core, parallel streams that operate in concert to address the specific Research Questions defined in the introduction section. The horizontal Information and Knowledge Mining Process addresses RQ1 and RQ2 by describing the value-adding lifecycle of a digital asset as it originates from a Producer and progresses from raw bits to a reasoned Knowledge Base available to a Consumer. The underlying Trustworthy Information and Knowledge Mining Long-Term Archiving Stream addresses RQ3 by ensuring the integrity, verifiability, and long-term preservation of the information and knowledge assets at every stage. This separation of concerns allows for the continuous enrichment of data without compromising its provenance, creating a virtuous cycle where data is not merely stored, but actively mined for intelligence.

The Information and Knowledge Mining Process represents the primary workflow for transforming raw inputs into actionable intelligence, structured as a linear progression through four distinct stages, each representing a continuous increase in semantic richness and analytical capability. The process begins with the Data stage, where the digital asset exists in its most primitive form: a raw Bitstream. This corresponds to context-free facts and signals, such as uninterpreted sensor logs or raw genomic sequences provided by a Producer, where the system has no understanding of what the asset represents beyond its binary existence. At this initial level, discovery is severely limited to Exact Match search; an asset can only be retrieved if the user possesses its exact filename or a bit-for-bit checksum. There is no insight into the content, effectively rendering the data invisible without external metadata. As the asset moves to the Content stage, the system parses its internal Structure and Format. This stage aligns with the capabilities of traditional Content Management Systems, where the system recognizes the asset as a specific file type, such as PDF or CSV, and can index its internal components. This structural awareness enables Index-based Syntactic Search, leveraging classic Information Retrieval techniques to perform full-text queries. However, the search remains purely syntactic, matching strings of characters rather than the concepts they represent, leaving the burden of interpretation entirely on the human user.

Moving forward to the third stage, and directly addressing the first research question regarding knowledge extraction, the asset is transformed into Knowledge by enriching it with a Semantic layer. This is the domain of Knowledge Extraction and Analysis, where AI-driven processes are applied to derive machine-readable meaning from the Content. This stage is critical because it provides the direct architectural solution to Remaining Challenge 1 (RC1): the effective integration of knowledge extracted from multimodal sources. The IKMF model solves this by treating the semantic layer as a universal integration point; whether an entity is extracted from a textual abstract, a table column, or an image caption, it is mapped to the same semantic concept. Consequently, discovery capabilities are elevated to Index-based Syntactic and Semantic Search, allowing users to query for concepts rather than just keywords, leveraging an understanding of meaning that aligns with modern dense retrieval methods.

The final stage of the mining process is the integration of these individual Knowledge assets into a formal Knowledge Base, adding the ultimate layer of Context. This addresses the second research question regarding formal representation. Here, extracted facts are synthesized into a cohesive model, such as a knowledge graph governed by a formal ontology. This constitutes the formal Designated Community Knowledge Base, making the implicit assumptions of a research community explicit and machine-actionable. Discovery at this stage transcends search and becomes Reasoning. A symbolic reasoner can utilize the formal rules of the domain to infer new, implicit knowledge from the explicitly asserted facts, delivering high-level insights to the Consumer. Furthermore, this continuous flow directly addresses Remaining Challenge 2 (RC2): managing the knowledge graph lifecycle. By defining the mining process as an ongoing, automated pipeline rather than a one-time ingestion event, the reference model provides a mechanism where the knowledge base is continuously updated and validated against the ontology as new data arrives.

Running in parallel and supporting every stage of this process is the Trustworthy Information and Knowledge Mining Long-Term Archiving Stream. Whereas the upper layer focuses on value creation, this stream focuses on value preservation and validity, ensuring that the system functions not just as a processor, but as a permanent record of scientific inquiry. It is composed of three continuous activities that underpin the model. Information and Knowledge Assembly involves collecting and connecting all representations of an asset throughout its lifecycle, combining the raw data stream, the parsed content, and the extracted knowledge annotations into a complete information package. This ensures a complete and auditable provenance trail for every insight, preventing the dissociation of results from their source data.

Crucially, the Information and Knowledge Reproducibility activity provides the direct solution to the third research question and Remaining Challenge 3 (RC3): preserving the computational reproducibility of scientific findings. While traditional archives preserve data, the IKMF model mandates the archiving of the computational context of the mining process itself. This means capturing the specific algorithms, software versions, dependencies, and parameter settings used to transform data into knowledge. By treating the analysis pipeline as an archival object, the system ensures that results are verifiable and repeatable, mitigating the black box problem often associated with AI-driven analysis. Finally, Information and Knowledge Preservation applies the rigorous principles of digital archiving, utilizing standards like OAIS and PREMIS, to protect these assembled packages against technological obsolescence. This holistic approach ensures that all assets, including data, content, knowledge, and the logic that connects them, remain Findable, Accessible, Interoperable, and Reusable for their Designated Community far into the future.

In conclusion, the IKMF Reference Model represents a meta-model for next-generation knowledge systems. It bridges the gap between static archiving and dynamic AI analysis, providing a blueprint for systems that not only manage data but actively participate in the scientific discovery process. The model systematically addresses the three core research questions by defining an Information and Knowledge Mining Process that extracts and structures knowledge (RQ1 and RQ2) and a parallel Trustworthy Long-Term Archiving Stream that ensures reproducibility and preservation (RQ3). This dual-stream approach directly resolves the identified remaining challenges of multimodal integration, knowledge graph lifecycle management, and computational reproducibility, laying the theoretical groundwork for the practical implementation and evaluation phases that follow.

\section{Discussion and Outlook}
The reference model for the IKMF, as detailed in the preceding sections, represents a direct and systematic response to the challenges of knowledge fragmentation that characterize contemporary data-intensive environments. The primary contribution of this work is not the invention of a single new software system, but rather the holistic architectural synthesis of multiple advanced technologies, including data integration, machine learning, semantic representation, and digital preservation, into a single, web-based, symbiotic ecosystem. By integrating the entire knowledge lifecycle, the IKMF aims to move beyond the limitations of our previous work and other piecemeal solutions. This integrated approach is designed to unlock the latent value in organizational and research data, which is often diminished when siloed, and to transform a passive data repository into an active partner in the scientific discovery process.

By systematically addressing research questions related to knowledge extraction (KEA), representation (KRR), and preservation (KAP), the IKMF provides a comprehensive roadmap for building next-generation knowledge management systems. A key strength of the design is its direct alignment with these research pillars. The KEA layer provides a scalable framework for converting raw data into structured information. The KAP layer, grounded in established standards like OAIS and integrating with systems like DSpace and a CRIS, ensures that knowledge assets are managed as permanent, trustworthy, and citable scholarly objects. The KRR layer, with its formal ontology, serves as the semantic core that transforms isolated facts into a coherent and computationally understandable knowledge base, enabling a far deeper level of inquiry than was previously possible.

A pivotal contribution of this proposed architecture is its explicit support for a hybrid, neuro-symbolic approach to AI. The IKMF moves beyond the dichotomy of choosing between traditional symbolic AI (e.g., reasoners) and modern generative AI (e.g., LLMs). Instead, it proposes a symbiotic framework, visualized in Figure \ref{fig:neuro_symbolic_cycle}, where the two paradigms work together with the enterprise ontology acting as the crucial bridge. This approach directly addresses the primary weaknesses of each technology. The inherent lack of verifiability and the risk of hallucination in LLM \cite{ji2023survey} are mitigated by using the symbolic deterministic reasoner as a real-time validation and integrity check engine. In contrast, the brittleness and labor-intensive knowledge acquisition process of symbolic systems are alleviated by using LLMs to accelerate knowledge extraction from unstructured text. This integration of sub-symbolic (neural) and symbolic reasoning is widely considered to be the "third wave" of AI, promising systems that are both flexible and trustworthy \cite{garcez2019neurosymbolic}. By designing a platform where the probabilistic outputs of LLMs can be grounded and verified against a formal, reproducible knowledge base, the IKMF provides a pragmatic roadmap for building knowledge systems that are not only intelligent but also accountable and reliable.

However, the realization of such a comprehensive vision is not without considerable challenges. The technical complexity of integrating these diverse components into a seamless, high-performance system is substantial. It requires careful definition of APIs and data exchange formats to avoid creating new internal silos. Furthermore, the effectiveness of the ecosystem is heavily dependent on semantic governance. The development and maintenance of the enterprise ontologies and taxonomies are not one-time tasks but require continuous effort from domain experts to ensure they remain accurate and relevant. From an organizational perspective, successful adoption requires a cultural shift towards data stewardship and collaborative knowledge sharing. Without sustained user participation and integration into existing workflows, even the most advanced system risks becoming an underutilized knowledge graveyard.

The work presented in this paper establishes the theoretical foundation required by the Design Science Research methodology \cite{hevner2004design}. The path forward involves a structured program of research and development. The immediate next step is the System Development phase, which will focus on translating the conceptual framework into a detailed technical architecture and implementing a minimum viable prototype. This will be followed by the Experimentation phase, involving a multi-stage evaluation that includes technical performance benchmarking, usability studies, and case studies with specific research groups to measure the system's impact on the quality and efficiency of knowledge discovery. The results will then feed back into iterative cycles of refinement. Ultimately, the vision for the IKMF is not to create a static product, but a foundational and evolving platform that can fundamentally alter how knowledge is created, managed, shared, and preserved in the 21st century.

\bibliographystyle{ieeetr}
\bibliography{references}

\end{document}